\documentstyle[twoside, 11pt]{article}

\newtheorem{propo}{Proposition}[section]

\newtheorem{theorem}{Theorem}

\newtheorem{lemma}{Lemma}[section]

\newtheorem{assumption}{Assumption}
\def\qed{ \ \vrule width.2cm height.2cm depth0cm\smallskip}
\def \wd {\wedge}
\def \bib {\bibitem}
\def \ind{1\!\!1}
\def \rw {\rightarrow}

\newcommand{\eps}{\varepsilon}

\newcommand{\brm}{\begin{rem}}
\newcommand{\ermq}{\end{rem}}

\newcommand{\ba}{\begin{array}}
\newcommand{\ea}{\end{array}}
\newcommand{\be}{\begin{equation}}
\newcommand{\ee}{\end{equation}}
\newcommand{\bea}{\begin{eqnarray}}
\newcommand{\eea}{\end{eqnarray}}
\newcommand{\beaa}{\begin{eqnarray*}}
\newcommand{\eeaa}{\end{eqnarray*}}

\def \R{I\!\!R}
\def \N{I\!\!N}

\def\a{\alpha}

\def\g{\gamma}

\def\t{\tau}

\def\G{\Gamma}

\def\cF{{\cal F}}

\def\cJ{{\cal J}}

\def\cT{{\cal T}}

\def\no{\noindent}

\def\ms{\medskip}
\def\bs{\bigskip}
\def\q{\quad}
\def\qq{\qquad}

\def\bF{{\bf F}}

\def\qed{ \hfill \vrule width.25cm height.25cm depth0cm\smallskip}

\newcommand{\basa}{\begin{assumption}}
\newcommand{\easa}{\end{assumption}}

\newcommand{\bas}{\begin{assum}}
\newcommand{\eas}{\end{assum}}

\def\liminf{\mathop{\underline{\rm lim}}}

\def\esssup{\mathop{\rm esssup}}

\def\dis{\displaystyle}

\def\bF{{\bf F}}

\newtheorem{thm}{Theorem}[section]

\newtheorem{cor}[thm]{Corollary}
\newtheorem{prop}[thm]{Proposition}
\newtheorem{rem}[thm]{Remark}

\newtheorem{defn}[thm]{Definition}
\newtheorem{assum}[thm]{Assumption}

\title{The Multi-player Nonzero-sum Dynkin Game in Continuous Time
}
\author{Said
Hamad\`ene\thanks{Universit\'e du Maine, LMM, Avenue Olivier
Messiaen, 72085 Le Mans, Cedex 9, France. e-mail:
hamadene@univ-lemans.fr}\,\,\,\,\,and \, Mohammed
Hassani\thanks{Universit\'e Cadi Ayyad, Facult\'e poly-disciplinaire
de Safi, D\'epartement de Math\'ematiques et Informatique. B.P. 4162
Safi Maroc. e-mail : medhassani@ucam.ac.ma.
  This work has been carried out while the second author was visiting Universit\'e du
Maine, Le Mans (Fr.).}}
\begin{document}
\date{\today}
\maketitle
\begin{abstract}In this paper we study the N-player nonzero-sum
Dynkin game ($N\geq 3$) in continuous time, which is a
non-cooperative game where the strategies are stopping times. We
show that the game has a Nash equilibrium point for general payoff
processes.
\end{abstract}
{\bf AMS Classification subjects}: 91A15 ; 91A10 ; 91A30 ; 60G40 ;
91A60.
\medskip

\no {$\bf Keywords$}: Nonzero-sum Game ; Dynkin game ; Snell
envelope ; Stopping time . $\qed$\ms
\section{Introduction}

A Dynkin game is a game where the controllers make use of stopping
times as control actions. Actually assume one has $N$ players
denoted by $\pi_1,...,\pi_N$ and each of which is allowed, according
to its advantages, to stop the evolution of a system. The system can
be for example an option contract which binds several agents
(players) in a financial market. So for $i=1,...,N$, assume that the
player $\pi_i$ makes the decision to stop the system at $\t_i$, then
its corresponding yield is given by: \be \label{eqpayoff}\ba{lll}
\dis J_i(\t_1,\cdots, \t_N) :={\bf E}[X^i_{\t_i}1_{\{\t_i<R_i \}} +
Q^i_{\t_i}1_{\{\t_i=R_i \}}+ Y^i_{R_i}1_{\{\t_i>R_i \}}]\ea \ee
where $R_i:=\min\{\t_j, j\neq i\}$ and $X^i, Q^i, Y^i$ are
stochastic processes described precisely below. This yield depends
actually on whether $\pi_i$ is the first to stop the evolution of
the system or not. So the main problem we are interested in is to
find a Nash equilibrium point (hereafter $NEP$ for short) for the
game, i.e., an $N$-uplet of stopping times $(\t_1^*,...,\t_N^*)$
such that for any $i=1,...,N$, for any $\t_i$,
$$ J_i(\t_1^*,\cdots, \t_N^*)\geq J_i(\t_1^*,...,\t_{i-1}^*,
\t_i,\t_{i+1}^*,...,\t_N^*).$$

A NEP is a collective strategy of stopping for the players which has
the feature that if one of them decides unilaterally to change a
strategy of stopping then it is penalized. \ms

In the case when $N=2$ and $J_1+J_2=0$, the game is called of {\it
zero-sum} type and the corresponding NEP is called a {\it
saddle-point} for the game. Otherwise it is called of {\it
nonzero-sum} type. \ms

The first works related to Dynkin games, which concern mainly the
zero-sum setting, go back to several decades before (see e.g.
\cite{bfried2, bismut, dynkin, fried, krylov1, krylov2, lepmaing,
stet, morimoto1, jz}, etc. and the references given there). This
latter setting was revisited several years later by many authors
especially in connection with reflected backward equations
\cite{CK}, the pricing of American game contingent claims introduced
by Y. Kifer (see e.g. \cite{kifer,hamadene,
 kkal, kuhn2}, etc.), convertible bonds (see e.g. \cite{crepey,shreve}) or other reasons (\cite{larakisolan,touzivielle}), etc.\ms

In comparison, nonzero-sum Dynkin games in continuous time, introduced by Bensoussan
and Friedman in \cite{bfried} have attracted few research activities (see e.g. \cite{bfried, catlep, etourneau,hz,
nagai}, etc.). Moreover, on the one hand, those papers deal with the
case of two players (even in \cite{bfried}) and, on the other hand,
the assumptions on the data of the problem are rather tough since,
except a recent paper by S.Hamad\`ene and J.Zhang \cite{hz}, authors
assume that e.g. the processes $Y^i$ of (\ref{eqpayoff}) are
supermartingales. Even in the discrete time setting, this latter assumption is supposed (e.g. in \cite{morimoto2}).

So the main objective of this paper is to study the continuous time nonzero-sum
Dynkin game in the case when there are more than two players and
general stochastic processes $X^i$, $Y^i$ and $Q^i$, $i=1,...,N$. The processes $Y^i$ are no 
longer supposed being supermartingales. We actually show this game has a Nash equilibrium point under minimal
assumptions. We should point out that according to our best knowledge
this problem has been not considered yet and the generalization from
the two-player setting to the multi-player one, even with respect to
the work by Hamad\`ene-Zhang \cite{hz}, is not formal and raises
questions which are far from to be obvious, especially the
construction of the approximating scheme and the study of its
properties.  \ms

This paper is organized as follows. In Section 2, we set accurately
the problem, recall the Snell envelope notion and provide a result
(Prop. \ref{streamline}) which is in a way the streamline in the
construction of the  of the NEP of the nonzero-sum Dynkin game. The
approximating scheme and its main properties are introduced in
Section 3. Finally in Section 4, we show that the limit of the
approximating scheme is a NEP for the game.   \qed
\section{Setting of the problem and hypotheses}
\setcounter{equation}{0} Throughout this paper $T$ is a positive real
constant which stands for the horizon of the problem and
$(\Omega ,\mathcal{F},P)$ is a
 fixed probability space on which is defined a filtration ${\bF}:=(\cF_{t})_{t\leq T}$
 which satisfies the usual conditions, $i.e.$, it is complete and right continuous.

 Next for any stopping time $\theta \leq T$, let us denote by:
 \medskip

(i) $\cT_\theta$ the set of $\bF$-stopping times $\tau$ such that P-a.s.
$\tau \in [\theta,T]$;

(ii) $E_\theta[.]$ the conditional expectation $w.r.t.$
$\cF_\theta$, i.e., ${\bf E}_\theta[X]:={\bf E}[X|{\cal F}_\theta]$,
for any integrable random variable $X$ ;

(iii) $\cJ:=\{1,...,N\}$.
\medskip

Next an $\bF$-progressively measurable $\R$-valued stochastic process $(\zeta_t)_{t\leq T}$
is called of class [D] if the set of random variables $\{\zeta_\tau,\,\,\tau \in {\cal T}_0\}$
is uniformly integrable. Now for  $i=1,...,
N$, let us introduce $\bF$-progressively measurable and $\R$-valued stochastic processes
of class [D], $X^i:=(X^i_t)_{t\leq T}$, $Q^i:=(Q^i_t)_{t\leq T}$ and $Y^i:=(Y^i_t)_{t\leq T}$, which moreover satisfy the following
hypotheses:
\begin{assum}
\label{assumption}${}$: For any $i=1,\dots,N$,
\medskip

\noindent $\bf {(A1)}$: $X^i$ is right continuous with left limits
(RCLL for short) and does not have negative predictable jumps ;
\medskip

\noindent $\bf {(A2)}$: The process $Y^i$ is right continuous, i.e.,
P-a.s., for any $t<T$, $Y^i_t=\lim_{s\searrow t}Y^i_s$ and of class
[D]. As a consequence if $(\gamma_n)_{n\geq 0}$ is a decreasing
sequence of $\bF$-stopping times then ${\bf E}[
Y^i_{\lim_n\gamma_n}]=\lim_n{\bf E}[ Y^i_{\gamma_n}]$ ;
\medskip

\noindent $\bf {(A3)}$: The processes $X^i$, $Y^i$ and $Q^i$ verify:
$$P-a.s.\,\,\forall t\leq T, X^i_t \le
Q^i_t\le
 Y^i_t;$$

\noindent $\bf {(A4)}$: For all $\tau\in \cT_0$, P-a.s. $$(Q^i_\tau <
Y^i_\tau ; \tau<T)\subseteq \bigcap_{j=1}^N (X^j_\tau < Y^j_\tau).$$

Note that assumptions (A1)-(A3) are minimal in order to solve the
problem, as for [A4], it is satisfied for e.g. for any $i\in \cJ$
and $\tau\in \cT_0$,  P-a.s. on $(\tau <T)$: $X_\tau^i<Y_\tau^i$
or\, $Q^i_\tau = Y^i_\tau$. \qed

\end{assum}

Next for $T_1, T_2,\cdots, T_N$ elements of $\cT_0$ and for $i\in
\cJ$, let us define $J_i(T_1,T_2,\cdots, T_N)$,  the payoff
associated with the player $i$, as follows: \be \label{J12} \ba{lll}
\dis J_i(T_1,T_2,\cdots, T_N) :={\bf E}\Big\{X^i_{T_i}1_{\{T_i<R_i
\}} + Q^i_{T_i}1_{\{T_i=R_i \}}+ Y^i_{R_i}1_{\{T_i>R_i \}} \Big\}.
\ea \ee where $R_i:=\min\{T_j, j\neq i\}=\wedge_{j=1,N; j\neq
i}T_j.$
\medskip

\noindent The meaning of those payoffs in this nonzero-sum Dynkin
game framework is the following: Assume that for any $i=1,...,N$,
the player $\pi_i$ makes the decision to use the stopping time $T_i$
as a strategy of stopping. Let $i_0\in {\cal J}$, then:

- $\pi_{i_0}$ will receive an amount equal to $X^{i_0}_{T_{i_0}}$ if
it decides unilaterally to stop controlling first. As for the other
players $\pi_j$, $j\neq i_0$, each one will receive an amount which
equals to $Y^j_{T_{i_0}}$;

- $\pi_{i_0}$ will receive an amount equal to $Q^{i_0}_{T_{i_0}}$ if
there is a commitment with one or more other players to stop first
the game. Each one of the players $j$ which do not get involved in
the commitment will receive $Y^j_{T_{i_0}}$.
\bigskip

We next precise the notion of equilibrium that we are looking for.
\begin{defn}
\label{equidefn} We say that $(T_1^*,T_2^*,\cdots, T_N^*)\in
{\cT_0}^N$ is a Nash equilibrium point of the Nonzero-sum Dynkin
game associated with $(J_i)_{i\in\cJ}$ if for all $i=1,\cdots,N$ and
all $ T_1,\cdots,T_N\in \cT_0$ we have:
 \be
\label{equilibrium} J_i(T_1^*,\cdots,T_{i-1}^*,T_i,T_{i+1}^*,\cdots,
T_N^*)\le J_i(T_1^*,\cdots,T_{i-1}^*,T_i^*,T_{i+1}^*,\cdots, T_N^*).
\ee
\end{defn}

The definition means that when the equilibrium is reached, is
penalized each one of the players which makes the decision to change
unilaterally its strategy of stopping. $\Box$
\bigskip

Next to begin with we give a result which in way is a streamline in
order to construct a NEP for the nonzero-sum Dynkin game. Actually
we have:

\begin{propo}\label{streamline}Assume there exist $N$ stopping times $(\t_i^*)_{i=1,N}$ and $N$ $\bF$-progressively measurable RCLL processes
 $(W^i)_{i=1,N}$, such that for any $i=1,...,N$, if we set $R_i=\t_1^* \wd ...\wd \t^*_{i-1}\wd \t^*_{i+1} \wd...\wd \t_N^*$ and
 $R=\t_1^* {\wd} ... \wd \t_N^*= \t^*_{i}\wd R_i$
  then:

(i) $(W^i_{t\wedge R})_{t\leq T}$ is an $\bF$-martingale and
$(W^i_{t\wedge R_i })_{t\leq T}$ is an $\bF$-supermatingale,

(ii) $\forall \,\,t\leq T$, $W^i_t\ind_{\{t<R_i\}}\geq X_t^i
\ind_{\{t<R_i]}$ and $W^i_{\t^*_i}\ind_{\{\t^*_i<R_i\}}=X^i_{\t^*_i}
\ind_{\{\t^*_i<R_i\}}$,

(iii) $W^i_{R_i}=Y^i_{R_i} \ind_{\{R_i<T\}}+Q^i_T\ind_{\{R_i=T\}}$
and
 $(Y^i_{R_i}-Q^i_{R_i})\ind_{\{R_i=\t_i^*<T\}}=0$.\\

Then the $N$-uplet of stopping times $(\t_i^*)_{i=1,N}$ is a Nash
equilibrium point for the nonzero-sum Dynkin game associated with
$(J_i)_{i\in \cJ}$. Moreover for any $i\in \cJ$,
$$J_i((\t_i^*)_{i=1,N})={\bf E}[W_0^i].$$
\end{propo}
{\bf Proof}: Actually for any $i\in \cJ$, since $(W^i_{t\wedge
R})_{t\leq T}$ is a martingale then
\begin{equation}\label{eq1}\begin{array}{ll}{\bf E}[W^i_0]&={\bf E}[W^i_{R}]\\
{}&={\bf E}[W^i_{\t^*_i}\ind_{\{\t^*_i<R_i\}}+
W^i_{R_i}\ind_{\{\t^*_i=R_i\}}+W^i_{R_i}\ind_{\{\t^*_i>R_i\}}].\end{array}\end{equation}
But taking into account the equalities of (iii) we get:
$$
\begin{array}{l}W^i_{R_i}\ind_{\{\t^*_i=R_i\}}+W^i_{R_i}\ind_{\{\t^*_i>R_i\}}
\\=(Y^i_{R_i} \ind_{\{R_i<T\}}+Q^i_T\ind_{\{R_i=T\}})\ind_{\{\t^*_i=R_i\}}+
(Y^i_{R_i} \ind_{\{R_i<T\}}+Q^i_T\ind_{\{R_i=T\}})\ind_{\{\t^*_i>R_i\}}\\
=Q^i_{R_i} \ind_{\{R_i=\t_i^*<T\}}+Q^i_T\ind_{\{R_i=\t_i^*=T\}}+
Y^i_{R_i} \ind_{\{R_i<\t_i^*\}}\\
=Q^i_{R_i} \ind_{\{R_i=\t_i^*\}}+Y^i_{R_i} \ind_{\{R_i<\t_i^*\}}
\end{array}$$
Making now the substitution in (\ref{eq1}) we deduce that:
$$\begin{array}{ll}{\bf E}[W^i_0]&=
{\bf E}[X^i_{\t^*_i} \ind_{\{\t^*_i<R_i\}}+Q^i_{\t^*_i}
\ind_{\{\t^*_i=R_i\}}+Y^i_{R_i} \ind_{\{\t^*_i>R_i\}}]
\\
{}&=J_i(\t_1^*,...,\t_N^*).\end{array}$$ On the other hand since
$(W^i_{t\wedge R_i })_{t\leq T}$ is an $\bF$-supermatingale then for
any stopping time $\g \in \cT_0$ we have,
\begin{equation}\label{eq3}\begin{array}{ll}{\bf E}[W^i_0]&\geq {\bf E}[W^i_{\g \wd R_i}]\\{}&={\bf E}[W^i_{\g}\ind_{\{\g<R_i\}}+
W^i_{R_i}\ind_{\{\g=R_i\}}+W^i_{R_i}\ind_{\{\g>R_i\}}].\end{array}\end{equation}
But once more
$$\label{eq2}\begin{array}{l}
W^i_{R_i}\ind_{\{\g=R_i\}}+W^i_{R_i}\ind_{\{\g>R_i\}}\\=
(Y^i_{R_i} \ind_{\{R_i<T\}}+Q^i_T\ind_{\{R_i=T\}})\ind_{\{\g=R_i\}}+(Y^i_{R_i} \ind_{\{R_i<T\}}+Q^i_T\ind_{\{R_i=T\}})\ind_{\{\g>R_i\}}\\
= (Y^i_{R_i} \ind_{\{R_i<T\}}+Q^i_T\ind_{\{R_i=T\}})\ind_{\{\g=R_i\}}+Y^i_{R_i} \ind_{\{\g>R_i\}}\\
\ge Q^i_{R_i}\ind_{\{\g=R_i\}}+Y^i_{R_i} \ind_{\{\g>R_i\}}
\end{array}
$$
since $Y^i\ge Q^i$. Plugging now this last term in (\ref{eq3}) and
since $W^i_{\g}\ind_{\{\g<R_i\}}\ge X^i_{\g}\ind_{\{\g<R_i\}}$ we
obtain:
$$\begin{array}{l}J_i(\t_1^*,...,\t_N^*)={\bf E}[W^i_0]\\\qq \geq
{\bf E}[X^i_{\g}\ind_{\{\g<R_i\}}+
Q^i_{R_i}\ind_{\{\g=R_i\}}+Y^i_{R_i}\ind_{\{\g>R_i\}}]=J_i(\t_1^*,...,\t_{i-1}^*,\g,\t_{i+1}^*,...,\t_N^*).\end{array}$$
Thus the $N$-uplet of stopping times $(\t_i^*)_{i=1,N}$ is a Nash
equilibrium for the nonzero-sum Dynkin game. \qed
\bigskip

To tackle the game problem, we mainly use the notion of Snell
envelope of processes which we introduce briefly now. For more
details on this subject one can refer e.g. to El-Karoui \cite{Elka}
or Dellacherie and Meyer \cite{DM}.
\begin{theorem}\label{snev} (\cite{DM}, pp. 431 or \cite{Elka}, pp. 140) \label{thmsnell}: Let $U=(U_t)_{0\le t\leq T}$ be an
$\bF$-adapted $\R$-valued RCLL process that belongs to class
[D]. Then there exists $Z:=(Z_t)_{0\le t\leq T}$ an
$\bF$-adapted $\R$-valued RCLL process of class [D],
such that $Z$ is the smallest super-martingale which dominates $U$,
$i.e$, if $(\bar{Z}_t)_{0\leq t\leq T}$ is another RCLL
supermartingale of class [D] such that $\bar{Z}\geq
U$ then $\bar{Z}\geq Z$, P-a.s.. The process
$Z$ is called the {\it Snell envelope} of $U$. It satisfies the following properties:

(i) For any $\bF$-stopping time $\theta$ we have:  \be \label{sun}
Z_\theta=\esssup_{\tau \in {\cal T}_{\theta}}{\bf
E}[U_\tau|\cF_\theta]\,\,\,\,\,\,(\mbox{and then }Z_T=U_T).\ee

(ii) If the predictable jumps of $U$ are only positive, then the
stopping time  $$\tau^*=\inf\{s\geq 0,\q Z_s=U_s\}$$ is optimal,
$i.e.$,
\begin{equation}\label{sdeux} {\bf E}[Z_0] ={\bf E}[Z_{\tau^*}]={\bf E}[U_{\tau^*}]=\sup_{\tau \geq 0}{\bf E}[U_\tau].\end{equation}
\begin{rem}\label{snelmart} As a by-product of (\ref{sdeux}) we have
$Z_{\tau^*}=U_{\tau^*}$ and the process $(Z_{t\wedge \tau^*})_{t\leq T} $ is a martingale. $\qed$\end{rem}
\end{theorem}
\section{The approximating scheme and its properties}

We are now going to introduce sequences of stopping times which, as
we will show it later, converge to a NEP of the game. So let us
consider the sequence of $\bF$-stopping times $ (\tau_n)_{n\geq1}$
defined, by induction, as follows: \bs \setcounter{equation}{0} \ms

\no (i) $\tau_1=\cdots\tau_N=T.$\\
(ii) For $n\geq N+1$, let $(i,q)=(i_n,q_n) \in \N ^2$ be such that
$n=Nq+i$ with $1\leq i\leq N$. Then let us set: $$
\begin{array}{ll}
-\q \theta_n= \min\{\tau_{n-1},\tau_{n-2},\cdots,\tau_{n-N+1}\}\,;\\
-\q  U^n_t= X^i_{t}1_{\{t<\theta_n \}} +
\widetilde{Y}^i_{\theta_n}1_{\{t\geq\theta_n \}}\q \hbox{ with }
\widetilde{Y}^i_t= Q^i_T1_{\{t=T \}}+Y^i_t1_{\{t<T \}},\,\,\forall
t\leq
T; \\
-\q \forall t\leq T,\,\,W^n_t=\esssup_{\nu \in {\cal T}_{t}}{\bf E}[U^n_\nu|\cF_t]\,;\\
-\q  \mu_n=\inf\{s\geq 0, W^n_s=U^n_s\}\, ;\\

-\q \tau_n=
(\mu_n\wedge\tau_{n-N})1_{\{\mu_n\wedge\tau_{n-N}<\theta_n\}}+\tau_{n-N}1_{\{\mu_n\wedge\tau_{n-N}\geq\theta_n\}}.

\end{array}$$

\begin{rem} \label{rem1}As a direct consequence of the above definitions and the properties of the Snell envelope,
the following relations or properties hold true: for any $n\geq N+1$,

\no (i) $W^n$ is RCLL and for any $t\geq\theta_n$,
$$
W^n_t=U^n_t=\widetilde{Y}^i_{\theta_n};$$ (ii) $$\mu_n \leq
\theta_n,\,\, \tau_n\leq\tau_{n-N}\mbox{ and }
\theta_{n}\leq\theta_{n-N};
$$
(iii) since the predictable jumps of $U^n$ are only positive and
taking into account Assumption (A3), we deduce from Theorem
\ref{snev}-(ii) that $\mu_n$ is optimal, i.e.,
$$
{\bf E}[W^n_0]={\bf E}[W^n_{\mu_n}]={\bf
E}[U^n_{\mu_n}]=\displaystyle\sup_{\tau \geq 0}{\bf E}[U^n_\tau].
\qq
$$
(iv) Let $n=Nq+i$ where the pair $(i,q)$ is as above, therefore even
if this is not explicitly mentioned in the definition, the stopping
time $\theta_{n}=\theta_{Nq+i}$ depends on $i$. The same happens for
$U^n$, $\t_n$, $\mu_n$ and $W^n$. $\Box$
\end{rem}

Additionally we have:

\begin{propo}
For any $n\geq 1$, $\mu_{n+N}\leq \t_n$, P-a.s..
\end{propo}
{\bf Proof}: Actually suppose there exists $m\geq 1$ such that
$P[\tau_m < \mu_{m+N}]>0$ and  let us set $n=\min\{m\geq1 \;
\hbox{s.t.}\; P[\tau_m < \mu_{m+N}]>0\}$. Then we obviously have
$n\geq N+1$. Next on the set $\{\tau_n < \mu_{n+N}\}$ we have:
\begin{equation}\label{eq4}\tau_n < \theta_{n+N}:= \tau_{n+N-1}\wedge
\tau_{n+N-2}\wedge\cdots \tau_{n+1}\end{equation}since
$\mu_{n+N}\leq \theta_{n+N}$ (Rem.\ref{rem1}-(ii)). But the
definition of $n$ implies that $\mu_{n+N-1}\leq \tau_{n-1}$ and then
$$\tau_{n+N-1}=\mu_{n+N-1} 1_{\{\mu_{n+N-1}
<\theta_{n+N-1}\}}+\tau_{n-1}1_{\{\mu_{n+N-1}=\theta_{n+N-1}\}}$$
and from (\ref{eq4}) and the definition of $\theta_{n+N-1}$ we
deduce that $\theta_{n+N-1}=\tau_n $. It follows that:
$$
\tau_{n+N-1}=\mu_{n+N-1} 1_{\{\mu_{n+N-1}
<\t_n\}}+\tau_{n-1}1_{\{\mu_{n+N-1}=\t_n\}}
$$Therefore
\be\label{eq5} \tau_n <\tau_{n+N-1} =\tau_{n-1}. \ee The strict
inequality stems from (\ref{eq4}) as for the equality it holds true
since $\mu_{n+N-1}\le \theta_{n+N-1}=\t_n$ and $\tau_n
<\tau_{n+N-1}$.

Next $$\theta_{n+N-2}:= \tau_{n+N-3}\wedge \tau_{n+N-4}\wedge\cdots
\tau_{n+1}\wd \t_n\wd \t_{n-1} =\t_n$$ since from (\ref{eq4}) for
any $k=1,...,N-1$ we have $\t_n<\t_{n+k}$ and from (\ref{eq5})
$\t_n<\t_{n-1}$. But once more the definitions of $n$ and
$\tau_{n+N-2}$ imply that:
$$
\tau_{n+N-2}=\mu_{n+N-2}
1_{\{\mu_{n+N-2}<\t_n\}}+\tau_{n-2}1_{\{\mu_{n+N-2}=\t_n\}}
$$
As we know that $\tau_{n+N-2}>\t_n$ then
$\tau_{n+N-2}=\tau_{n-2}>\t_n$.

Repeating now this procedure as many times as necessary we deduce
that for any $j=1,\dots, N-1$,
$$
\t_n<\t_{n+N-j}=\t_{n-j}$$ and then on the set $\{\t_n<\mu_{n+N}\}$
we have
$$
\t_n<\theta_{n+N}=\theta_n$$ and thanks to the definitions of $n$
and $\t_n$ we also have on $\Gamma:=\{\t_n<\mu_{n+N}\}$
$$\tau_n=\mu_{n} 1_{\{\mu_{n}
<\theta_{n}\}}+\tau_{n-N}1_{\{\mu_{n}=\theta_{n}\}}=\mu_{n}
$$ since $\mu_n\leq \t_{n-N}$. Therefore on the set $\Gamma\in \cF_{\t_n}$ we
have $U^n=U^{n+N}$ since $\theta_{n+N}=\theta_n$ and
$$\begin{array}{lll} \ind_\G W^{n+N}_{\mu_n}=\ind_\G
W^{n+N}_{\t_n}&=&\ind_\G\esssup_{\nu \in {\cal T}_{\t_n}}{\bf E}[
U^{n+N}_\nu|\cF_{\t_n}]=\esssup_{\nu \in {\cal T}_{\t_n}}{\bf
E}[\ind_\G U^{n+N}_\nu|\cF_{\t_n}]
\\ & =& \esssup_{\nu \in {\cal T}_{\t_n}}{\bf E}[\ind_\G U^{n}_\nu|\cF_{\t_n}]
\\ &=& \ind_\G W^{n}_{\t_n}=\ind_\G W^{n}_{\mu_n}
 \\ &=& \ind_\G U^{n}_{\mu_n}
 \\ &=& \ind_\G U^{n+N}_{\mu_n},
\end{array}
$$ i.e., $\ind_\G W^{n+N}_{\mu_n}=\ind_\G U^{n+N}_{\mu_n}$ and then
$\mu_{n+N}\leq \mu_n$ on $\G$ since $\mu_{n+N}$ is the first time
that $W^{n+N}$ reaches $U^{n+N}$. As on $\G$ we have
$\mu_n=\t_n<\mu_{n+N}$ then this is contradictory with the previous
inequality. It follows that $P[\G]=0$ and for any $m\geq 1$ we have
$\mu_{m+N}\leq \t_m$, P-a.s.. The proof is complete. \qed \ms

As a by-product we obtain:

\begin{cor}\label{cor1} For any $n\geq N+1$,

(i) $ \tau_n=
\mu_n1_{\{\mu_n<\theta_n\}}+\tau_{n-N}1_{\{\mu_n=\theta_n\}}$ ;

(ii)
$\mu_n=\tau_n\wedge\theta_n=\tau_n\wedge\tau_{n-1}\wedge\cdots\tau_{n-N+1}
.$
\end{cor}
{\bf Proof}: Indeed, (i) is a direct consequence of the previous
proposition and the definition of $\t_n$. As for (ii), we have
$$\ba{ll} \tau_n\wedge\theta_n&=\tau_n \ind_{\{\tau_n< \theta_n\}}+
\theta_n \ind_{\{\tau_n\geq \theta_n\}}.\ea$$But $\tau_n
\ind_{\{\tau_n< \theta_n\}}=\mu_n \ind_{\{\mu_n< \theta_n\}} $ and
on $[\tau_n\geq \theta_n]$ we have $\theta_n=\mu_n$. Therefore
$\theta_n \ind_{\{\tau_n\geq \theta_n\}}=\mu_n \ind_{\{\t_n\geq
\theta_n\}} $. Gathering now those relations yields
$\mu_n=\tau_n\wedge\theta_n$. Finally the second equality is just
the definition of $\theta_n$. \qed
\section{Existence of a Nash equilibrium point}

For any $i\in \{1,\cdots,N\}$, let us define:
$$T_i^*=\displaystyle\lim_{n\longrightarrow\infty}\;\tau_{Nn+i}\; \hbox{
and }\;
R^*_i=\displaystyle\lim_{n\longrightarrow\infty}\;\theta_{Nn+i}=\min\{T_j^*
\; ; j\neq i\}.$$ Those limits exist since for any $n\geq N+1$, we
know that $\t_n\leq \t_{n-N}$ therefore the sequences
$(\tau_{Nn+i})_{n\geq 0}$ are non-increasing for fixed $i$. \\

We have also, for all $i\in \{1,\cdots,N\}$
$$R^*:=T_1^*\wedge\cdots\wedge
T_N^*=\displaystyle\lim_{n\longrightarrow\infty}\;\mu_{Nn+i}
=\displaystyle\lim_{n\longrightarrow\infty}\;\mu_n. \Box$$\ms

We are going now to show that the $N$-uplet of stopping times
$(T_i^*)_{i=1,...,N}$ is a Nash equilibrium point for the N-players
nonzero-sum Dynkin game associated with $(J_i)_{i\in \cJ}$. The
proof will be obtained after several intermediary results given
below. \bs

\begin{lemma}\label{lem1}
For any $n\geq1$, $i\in\{1,\cdots,N\}$ and any $\theta\in {\cal
T}_{0}$ we have:
$$\begin{array}{ll}
J_i(\tau_{Nn+N+1},\tau_{Nn+N+2},\cdots,\tau_{Nn+N+i-1},\theta,\tau_{Nn+i+1},\cdots,
\tau_{Nn+N})
\\
\qquad \leq
J_i(\tau_{Nn+N+1},\tau_{Nn+N+2},\cdots,\tau_{Nn+N+i-1},\tau_{Nn+N+i},
\tau_{Nn+i+1},\cdots,\tau_{Nn+N})\\
\qquad \qq \qq + {\bf
E}[(Y^i_{\tau_{Nn+N+i}}-Q^i_{\tau_{Nn+N+i}})\;\ind_{\{\tau_{Nn+N+i}=\theta_{Nn+N+i}<T\}}].
\end{array}$$
\end{lemma}
\noindent {\bf Proof}: For any $n\geq1$, $i\in\{1,\cdots,N\}$ and
$\theta\in {\cal T}_{0}$,
$$\begin{array}{ll}
J_i(\tau_{Nn+N+1},\tau_{Nn+N+2},\cdots,\tau_{Nn+N+i-1},\theta,\tau_{Nn+i+1},
\cdots,\tau_{Nn+N})\\\qq\qq=E\Big\{X^i_{\theta}\ind_{\{\theta<\theta_{Nn+N+i}
\}} + Q^i_{\theta}\ind_{\{\theta= \theta_{Nn+N+i}\}}+
Y^i_{\theta_{Nn+N+i}}\ind_{\{\theta>\theta_{Nn+N+i} \}} \Big\}.
\end{array}
$$
But
$$X^i_{\theta}\ind_{\{\theta<\theta_{Nn+N+i}
\}}\le W^{Nn+N+i}_{\theta}\ind_{\{\theta<\theta_{Nn+N+i} \}}
$$
and since $Q^i\leq Y^i$ we have
$$\ba{l}
Q^i_{\theta}\ind_{\{\theta= \theta_{Nn+N+i}\}}+
Y^i_{\theta_{Nn+N+i}}\ind_{\{\theta>\theta_{Nn+N+i} \}}\\
\qq\qq\qq \leq(Q^i_{T}\ind_{\{\theta_{Nn+N+i}=T\}}+
Y^i_{\theta_{Nn+N+i}}\ind_{\{T>\theta_{Nn+N+i} \}})\ind_{\{\theta
\geq \theta_{Nn+N+i}\}}\\ \qq\qq\qq  \leq
W^{Nn+N+i}_{\theta_{Nn+N+i}}\ind_{\{\theta \geq \theta_{Nn+N+i}\}}.
\ea
$$
Therefore \be \label{eq6}\ba{l}
J_i(\tau_{Nn+N+1},\tau_{Nn+N+2},\cdots,\tau_{Nn+N+i-1},\theta,\tau_{Nn+i+1},
\cdots,\tau_{Nn+N})\\\qq\qq\qq \qq \leq  {\bf E}[
W^{Nn+N+i}_{\theta\wedge\theta_{Nn+N+i}}]\leq {\bf
E}[W^{Nn+N+i}_{0}]\ea \ee since $W^{Nn+N+i}$ is a supermartingale.

Next \be\label{eq7}\begin{array}{ll}
J_i(\tau_{Nn+N+1},\tau_{Nn+N+2},\cdots,\tau_{Nn+N+i-1},\tau_{Nn+N+i},\tau_{Nn+i+1},\cdots,\tau_{Nn+N})
\\
\qquad = {\bf E}[ U^{Nn+N+i}_{\tau_{Nn+N+i}\wedge\theta_{Nn+N+i}} ]+
 {\bf E}[(Q^i_{\theta_{Nn+N+i}} - Y^i_{\theta_{Nn+N+i}})\;
\ind_{(\tau_{Nn+N+i}=\theta_{Nn+N+i}<T)}] \\
\qquad = E[ U^{Nn+N+i}_{\mu_{Nn+i+1}} ]+
 {\bf E}[(Q^i_{\theta_{Nn+N+i}} - Y^i_{\theta_{Nn+N+i}})\;
\ind_{\{\tau_{Nn+N+i}=\theta_{Nn+N+i}<T\}}]\\
\qquad = E[ W^{Nn+N+i}_{\mu_{Nn+i+1}} ]+
 {\bf E}[(Q^i_{\theta_{Nn+N+i}} - Y^i_{\theta_{Nn+N+i}})\;
\ind_{\{\tau_{Nn+N+i}=\theta_{Nn+N+i}<T\}}]\\\qquad  = {\bf
E}[W^{Nn+N+i}_0]+ {\bf E}[(Q^i_{\theta_{Nn+N+i}} -
Y^i_{\theta_{Nn+N+i}})\; \ind_{\{\tau_{Nn+N+i}=\theta_{Nn+N+i}<T\}}]
\end{array}
\ee since $\tau_{Nn+N+i}\wedge\theta_{Nn+N+i}=\mu_{Nn+i+1}$ (Cor.
\ref{cor1}-(ii)), $(W^{Nn+N+i}_{t\wd \mu_{Nn+i+1}})_{t\leq T}$ is a
martingale (Remark \ref{snelmart}) and finally by (i) of Remark \ref{rem1}. Comparing now (\ref{eq6}) and (\ref{eq7}) to obtain the
desired result. \qed \ms

We now focus on the limits of the terms that appear in the
inequality of the previous lemma.

\begin{lemma}\label{lem2}: The following asymptotic inequalities hold true:

\no (i) For all  $\theta\in {\cal T}_{0}$ and $i\in \cJ$,  we have:
$$\begin{array}{ll}
\lim_{n\rightarrow \infty}
J_i(\tau_{Nn+N+1},\tau_{Nn+N+2},\cdots,\tau_{Nn+N+i-1},\theta,
\tau_{Nn+i+1},\cdots,\tau_{Nn+N})=
\\
\qquad
J_i(T^*_{1},T^*_{2},\cdots,T^*_{i-1},\theta,T^*_{i+1},\cdots,T^*_{N})-
{\bf
E}[(Q^i_{\theta}-X^i_{\theta})\;\ind_{\displaystyle\bigcap_{n\geq0}\{\theta=R^*_{i}
<\theta_{Nn+i}\}}].
\end{array}$$
(ii)
$$\begin{array}{ll}
\liminf_{n\rightarrow \infty} \bigg (
 J_i(\tau_{Nn+N+1},\tau_{Nn+N+2},\cdots,\tau_{Nn+N+i-1},\tau_{Nn+N+i},\tau_{Nn+i+1},\cdots,\tau_{Nn+N})+\\
\qquad \qq \qq {\bf E}[(Y^i_{\tau_{Nn+N+i}}-Q^i_{\tau_{Nn+N+i}})\;
\ind_{\{\tau_{Nn+N+i}=\theta_{Nn+N+i}<T\}}]\bigg) \; = \\
J_i(T^*_{1},T^*_{2},\cdots,T^*_{i-1},T^*_{i},T^*_{i+1},\cdots,T^*_{N})+
{\bf E}[ (Y^i_{T^*_{i}}-Q^i_{T^*_{i}})\; \ind_{\{T^*_{i}=R^*_{i}<T\}} ] + \\
\qq\qq \qq\qq \liminf_{n\rightarrow \infty} {\bf E}[(
X^i_{T^*_{i}}-Y^i_{T^*_{i}})\; \ind_{\{
\tau_{Nn+N+i}<\theta_{Nn+N+i},\;T^*_{i}=R^*_{i}<T\}}].
\end{array}$$
\end{lemma}
\noindent {\bf Proof}: (i) Actually \be\label{eq8}\ba{ll}
&J_i(\tau_{Nn+N+1},\tau_{Nn+N+2},\cdots,\tau_{Nn+N+i-1},\theta,
\tau_{Nn+i+1},\cdots,\tau_{Nn+N})\\{}&\qq\qq={\bf E}[X^i_\theta
\ind_{\{\theta\leq \theta_{Nn+N+i}\}}+Y^i_{\theta_{Nn+N+i}}
\ind_{\{\theta>\theta_{Nn+N+i}\}}+(Q^i_\theta-X^i_\theta)\ind_{\{\theta=
\theta_{Nn+N+i}\}}].\ea\ee As the process $Y^i$ is RCLL and of class
[D] and the sequence $(\theta_{Nn+N+i})_{n}$ is decreasing then,
when $n\rightarrow \infty$, \be \label{lim1}{\bf E}[X^i_\theta
\ind_{\{\theta\leq \theta_{Nn+N+i}\}}+Y^i_{\theta_{Nn+N+i}}
\ind_{\{\theta>\theta_{Nn+N+i}\}}]\rightarrow {\bf E}[X^i_\theta
\ind_{\{\theta\leq R_i^*\}}+Y^i_{R_i^*} \ind_{\{\theta>R_i^*\}}].
\ee On the other hand, when $n\rightarrow \infty$, \be
\label{lim2}{\bf E}[(Q^i_\theta-X^i_\theta)\ind_{\{\theta=
\theta_{Nn+N+i}\}}]\rightarrow {\bf
E}[(Q^i_\theta-X^i_\theta)\ind_{\{\theta= R_i^*\}}]- {\bf
E}[(Q^i_{\theta}-X^i_{\theta})\;\ind_{\displaystyle\bigcap_{n\geq0}(\theta=R^*_{i}
<\theta_{Nn+i})}].\ee Actually (\ref{lim2}) is obtained in paying
attention whether the sequence $(\theta_{Nn+N+i})_n$ is of
stationary type or not. Going back now to (\ref{eq8}), take the
limit and make use of (\ref{lim1}) and (\ref{lim2}) to obtain the
desired result . $\Box$ \ms

Next let us focus on (ii). Let $i\in \cJ$ be fixed, then:
$$\begin{array}{ll} \liminf_{n\rightarrow \infty} \bigg (
 J_i(\tau_{Nn+N+1},\tau_{Nn+N+2},\cdots,\tau_{Nn+N+i-1},
 \tau_{Nn+N+i},\tau_{Nn+i+1},\cdots,\tau_{Nn+N})+\\
  \qquad \qq\qq\qq {\bf E}[(Y^i_{\tau_{Nn+N+i}}-Q^i_{\tau_{Nn+N+i}})\;\ind_{\{\tau_{Nn+N+i}=\theta_{Nn+N+i}<T\}}]\bigg)
\;  \\ \\ = \liminf_{n\rightarrow \infty} E[X^i_{\tau_{Nn+N+i}}\;
\ind_{\{ \tau_{Nn+N+i}<\theta_{Nn+N+i}\}}
+Y^i_{\theta_{Nn+N+i}}\; \ind_{\{ \tau_{Nn+N+i}>\theta_{Nn+N+i}\}}+\\
\qquad \qq\qq\qq Y^i_{\theta_{Nn+N+i}}\; \ind_{\{
\tau_{Nn+N+i}=\theta_{Nn+N+i}<T\}}+Q^i_{T}\; \ind_{\{
\tau_{Nn+N+i}=\theta_{Nn+N+i}=T\}}] \;\\ \\
  = \liminf_{n\rightarrow \infty} E[X^i_{{\tau_{Nn+N+i}}}\; 1_{\{
\tau_{Nn+N+i}<\theta_{Nn+N+i};\;T^*_i\leq R^*_i\}}
+Y^i_{\theta_{Nn+N+i}}\; \ind_{\{ \tau_{Nn+N+i}>\theta_{Nn+N+i};\;T^*_i\geq R^*_i\}}\\
\qquad \qq \qq\qq\qq +Y^i_{\theta_{Nn+N+i}}\; \ind_{\{
\tau_{Nn+N+i}=\theta_{Nn+N+i};\;T^*_i=R^*_i<T\}}+Q^i_{T}\; 1_{\{
T^*_i=R^*_i=T\}}]  \; \\ \\ =
  E[X^i_{T^*_i}\; \ind_{\{
T^*_i< R^*_i\}}+ Q^i_{T^*_i}\; \ind_{\{ T^*_i=R^*_i\}}
+Y^i_{R^*_i}\; 1_{\{ T^*_i> R^*_i\}}]-{\bf E}[
Q^i_{T^*_i}\; \ind_{\{ T^*_i=R^*_i<T\}}]+\\
 \qq\qq\qq \liminf_{n\rightarrow \infty} E[X^i_{T^*_i}\; \ind_{\{
\tau_{Nn+N+i}<\theta_{Nn+N+i};\;T^*_i= R^*_i\}} +Y^i_{R^*_i}\;
\ind_{\{ \tau_{Nn+N+i}\geq\theta_{Nn+N+i};\;T^*_i= R^*_i<T\}}]\\
\\=
J_i(T^*_{1},T^*_{2},\cdots,T^*_{i-1},T^*_{i},T^*_{i+1},\cdots,T^*_{N})+
E [(Y^i_{T^*_{i}}-Q^i_{T^*_{i}})\; 1_{\{T^*_{i}=R^*_{i}<T\}} ] + \\
\qq\qq\qq \qq\qq\qq \liminf_{n\rightarrow \infty} {\bf E}[(
X^i_{T^*_{i}}-Y^i_{T^*_{i}})\; \ind_{\{
\tau_{Nn+N+i}<\theta_{Nn+N+i},\;T^*_{i}=R^*_{i}<T\}}]
\end{array}$$
which is the desired result. Note that in the fourth inequality we
have taken into account the fact that the processes $X^i$ and $Y^i$
are of RCLL and of class [D]. \qed \bs

An obvious consequence of Lemmata \ref{lem1} and \ref{lem2} is:
\begin{cor}\label{cor1}
For all  $\theta\in {\cal T}_{0}$ and all $i\in\cJ$
\be\label{ineqcor}\begin{array}{ll}
J_i(T^*_{1},T^*_{2},\cdots,T^*_{i-1},\theta,T^*_{i+1},\cdots,T^*_{N})-
{\bf E}[(Q^i_{\theta}-X^i_{\theta})\;
\ind_{\displaystyle\bigcap_{n\geq0}\{\theta=R^*_{i}<\theta_{Nn+i}\}}]\\\qq
\leq
J_i(T^*_{1},T^*_{2},\cdots,T^*_{i-1},T^*_{i},T^*_{i+1},\cdots,T^*_{N})+
E [(Y^i_{T^*_{i}}-Q^i_{T^*_{i}})\; \ind_{\{T^*_{i}=R^*_{i}<T\}}]  \\
\qq \qq \qq \qq \qq +\liminf_{n\rightarrow \infty} {\bf E}[(
X^i_{T^*_{i}}-Y^i_{T^*_{i}})\; \ind_{\{
\tau_{Nn+N+i}<\theta_{Nn+N+i},\;T^*_{i}=R^*_{i}<T\}}] .\qq \qed
\end{array}\ee
\end{cor}

\begin{lemma}\label{lem3} (i) We have:
$$\lim_{n \rightarrow \infty}{\bf E}[( X^i_{T^*_{i}}-Y^i_{T^*_{i}})\; \ind_{\{
\tau_{Nn+N+i}<\theta_{Nn+N+i},\;T^*_{i}=R^*_{i}<T\}}]=0$$and then
for all $\varepsilon>0$
$$\lim_{n \rightarrow \infty} P[Y^i_{T^*_{i}}-X^i_{T^*_{i}}>\varepsilon,\;\;
\tau_{Nn+N+i}<\theta_{Nn+N+i},\;\;T^*_{i}=R^*_{i}<T]=0.$$

(ii) For all  $\theta\in {\cal T}_{0}$ and all $i\in\cJ$,
$$\begin{array}{ll}
J_i(T^*_{1},T^*_{2},\cdots,T^*_{i-1},\theta,T^*_{i+1},\cdots,T^*_{N})+
E [(Y^i_{R^*_{i}}-Q^i_{R^*_{i}})\; \ind_{\{\theta =R^*_{i}<T\}}]
\\\qq\qq\qq\leq
J_i(T^*_{1},T^*_{2},\cdots,T^*_{i-1},T^*_{i},T^*_{i+1},\cdots,T^*_{N})+
{\bf E}[ (Y^i_{R^*_{i}}-Q^i_{R^*_{i}})\;
\ind_{\{T^*_{i}=R^*_{i}<T\}}].
\end{array}$$
\end{lemma}
\noindent {\bf Proof}: (i) Actually let $\theta$ be the following
$\bF$-stopping time: $$\theta=T^*_{i}\;\ind_{\{T^*_{i}<R^*_{i}\}} +
T\;1_{\{T^*_{i}\geq R^*_{i}\}}.$$ Then using inequality
(\ref{ineqcor}) yields:
$$\begin{array}{ll}
J_i(T^*_{1},T^*_{2},\cdots,T^*_{i-1},\theta,T^*_{i+1},\cdots,T^*_{N})-
{\bf E}[(Q^i_{\theta}-X^i_{\theta})\;
\ind_{\displaystyle\bigcap_{n\geq0}\{\theta=R^*_{i}<\theta_{Nn+i}\}}]
\\
\qq
=J_i(T^*_{1},T^*_{2},\cdots,T^*_{i-1},T^*_{i},T^*_{i+1},\cdots,T^*_{N})+
E [(Y^i_{T^*_{i}}-Q^i_{T^*_{i}})\; \ind_{\{T^*_{i}=R^*_{i}<T\}}]
\\ \qq \leq
J_i(T^*_{1},T^*_{2},\cdots,T^*_{i-1},T^*_{i},T^*_{i+1},\cdots,T^*_{N})+
{\bf E}[ (Y^i_{T^*_{i}}-Q^i_{T^*_{i}})\;
\ind_{\{T^*_{i}=R^*_{i}<T\}} ]
\\\qq \qq\qq + \liminf_{n\rightarrow \infty} {\bf E}[(
X^i_{T^*_{i}}-Y^i_{T^*_{i}})\; \ind_{\{
\tau_{Nn+N+i}<\theta_{Nn+N+i},\;T^*_{i}=R^*_{i}<T\}}] .
\end{array}$$
Hence $$\liminf_{n\rw \infty} {\bf E}[( X^i_{T^*_{i}}-Y^i_{T^*_{i}})\;
\ind_{\{ \tau_{Nn+N+i}<\theta_{Nn+N+i},\;T^*_{i}=R^*_{i}<T\}}] \geq
0
$$ which completes the proof since $X^i_{T^*_{i}}-Y^i_{T^*_{i}}\leq
0$ thanks to Assumption (A3). \ms

(ii) Let $\widetilde{\theta}$ be the following $\bF$-stopping
time:$$\widetilde{\theta}=\theta\; \ind_{\{\theta=R^*_{i}<T)^c} +
T\;\ind_{\{\theta=R^*_{i}<T\}}$$ where the superscript  $(^c)$
stands for the complement. Since $P[\tilde \theta =R^*_i<T]=0$ we
obtain from (\ref{ineqcor}) and (i),
$$\begin{array}{ll}
J_i(T^*_{1},T^*_{2},\cdots,T^*_{i-1},\widetilde{\theta},T^*_{i+1},\cdots,T^*_{N})-
{\bf E}[(Q^i_{\widetilde{\theta}}-X^i_{\widetilde{\theta}})\;
\ind_{\displaystyle\bigcap_{n\geq0}\{\widetilde{\theta}=R^*_{i}<\theta_{Nn+i}\}}]
\\\qq =
J_i(T^*_{1},T^*_{2},\cdots,T^*_{i-1},\theta,T^*_{i+1},\cdots,T^*_{N})+
E [(Y^i_{R^*_{i}}-Q^i_{R^*_{i}})\; \ind_{\{\theta =R^*_{i}<T\}}]
\\\qq\leq
J_i(T^*_{1},T^*_{2},\cdots,T^*_{i-1},T^*_{i},T^*_{i+1},\cdots,T^*_{N})+
E [(Y^i_{R^*_{i}}-Q^i_{R^*_{i}})\; \ind_{\{T^*_{i}=R^*_{i}<T\}}],
\end{array}$$
whence the desired result. \qed \ms

We now give a key-result which allows us to conclude.
\begin{propo}\label{prop1}
Under Assumption (A4),  for all $i\in \cJ$ we have: \be
\label{nul5}E [(Y^i_{T^*_{i}}-Q^i_{T^*_{i}})\;
\ind_{\{T^*_{i}=R^*_{i}<T\}}]=0.\ee
\end{propo}
\noindent {\bf Proof}: First note that for any $i\in \cJ$ we have:
\\
$$\begin{array}{l}
{\bf E}[(Y^i_{T^*_{i}}-Q^i_{T^*_{i}})
\ind_{\{T^*_{i}=R^*_{i}<T\}}]\leq\\\qq\qq \qq\qq
\displaystyle{\sum_{\tiny{I=\{i_1,...,i_k\} \subset \cJ,\,\, i\in I
\mbox{ and } k\geq 2}}\!\!\!\!\!\! \!\!\!\!\!\! E
[(Y^i_{T^*_{i}}-Q^i_{T^*_{i}})
\ind_{\{T^*_{i_1}=T^*_{i_2}=...=T^*_{i_k}<R^*_{I}\}}\}},
\end{array}$$\\
where $R^*_{I}= \min\{T^*_{j}\; ;\; j\notin I \}$ with $\min
\emptyset = T$. Therefore it is enough to show that for any
$i_1,...,i_k \in \cJ$, which we assume w.l.o.g satisfying
$i_1<i_2<\cdots<i_k$, we have:
$$
E [(Y^i_{T^*_{i}}-Q^i_{T^*_{i}})
\ind_{\{T^*_{i_1}=T^*_{i_2}=\cdots=T^*_{i_k}<R^*_{I}\}}]=0$$ for any
$i\in I=\{i_1,\cdots,i_k\}$. \ms

\no {\bf Step 1}: For any $n\ge 0$, \be \label{nul4}
P[A_n:=\bigcap_{j=1}^k\; \{
R^*_{I}>\tau_{Nn+i_j}\geq\theta_{Nn+i_j}\}]=0.\ee

Actually, first note that $P[A_0]=P[A_1]=0$. Next let us show that
$A_n\subset A_{n-1}$ for any $n\geq 2$.

 On $A_n$:

By the definitions of $\t_n$ and $R^*_I$, we have: $\forall j\in
\{1,\cdots,k\}$, $\forall \alpha\notin I$, \be \label{nul}
\tau_{Nn+i_j}=\tau_{N(n-1)+i_j} \;\hbox{  and }\;
\tau_{Nn+i_j}<R^*_{I}\leq\tau_{Nn+\alpha}\le \t_{N(n-\ell)+\a},
\,\,\ell \geq 0.\ee Therefore in using those properties we deduce
that: \be \label{nul2}\begin{array}{ll} &\theta_{Nn+i_j} :=
\tau_{Nn+i_j-1}\wedge\tau_{Nn+i_j-2}\wedge\cdots\wedge\tau_{Nn+i_j-N+1}\\
&\q
=\tau_{Nn+i_{j-1}}\wedge\tau_{Nn+i_{j-2}}\wedge\cdots\wedge\tau_{Nn+i_{1}}\wedge\tau_{Nn+i_{k}-N}
\wedge\tau_{Nn+i_{k-1}-N}\wedge\cdots\wedge\tau_{Nn+i_{j+1}-N}
\end{array} \ee
Let us give briefly the justification of the second equality. Indeed
for some $\ell \in \{1,...,N-1\}$ either $i_j-\ell >0$ or $i_j-\ell
\le 0$. Case (i): $i_j-\ell >0$. Then if $i_j-\ell \in
\{i_1,...,i_{j-1}\}$ then we keep it in the expression of
$\theta_{Nn+i_j}$ and if $i_j-\ell \notin \{i_1,...,i_{j-1}\}$ then
we know from (\ref{nul}) that, e.g., $\tau_{Nn+i_j-\ell}\ge
\tau_{Nn+i_{j-1}}$ and then $\tau_{Nn+i_j-\ell}$ is deleted from the
expression of $\theta_{Nn+i_j}$. Case (ii): $i_j-\ell \leq 0$. Then
$\tau_{Nn+i_j-\ell}=\tau_{N(n-1)+i_j-\ell+N}$ with $i_j-\ell+N\ge
i_j+1$. Once more if $i_j-\ell+N\in \{i_{j+1},...,i_{k}\}$ then we
keep $\tau_{Nn+i_j-\ell}$ it in the expression of $\theta_{Nn+i_j}$.
Otherwise, i.e., if $i_j-\ell+N\notin \{i_{j+1},...,i_{k}\}$, then
$\tau_{Nn+i_j-\ell}=\tau_{N(n-1)+i_j-\ell+N}\geq
\tau_{Nn+i_j-\ell+N}\geq \tau_{Nn+i_{j-1}}$ and $\tau_{Nn+i_j-\ell}$
is deleted from the expression of $\theta_{Nn+i_j}$. Thus we are
done. \ms

Now the first equality of (\ref{nul}) yields: \be\label{nul3}
\theta_{Nn+i_j} =
\tau_{Nn+i_{j-1}}\wedge\tau_{Nn+i_{j-2}}\wedge\cdots\wedge\tau_{Nn+i_{1}}\wedge\tau_{Nn+i_{k}}
\wedge\tau_{Nn+i_{k-1}}\wedge\cdots\wedge\tau_{Nn+i_{j+1}}.\ee

Next by a backward induction argument we have that for any $j\in
\{1,\cdots,k\}$,
$$\tau_{Nn+i_j}=\tau_{N(n-1)+i_j}=\tau_{N(n-2)+i_j} \mbox{ and
}\theta_{Nn+i_j}=\theta_{N(n-1)+i_j}.
$$
Actually for $j=k$, by (\ref{nul}) and (\ref{nul3}) we have:
$$\begin{array}{ll}
 \tau_{Nn+i_k}=\tau_{N(n-1)+i_k}\geq\theta_{Nn+i_k}
&
=\tau_{Nn+i_{k-1}}\wedge\tau_{Nn+i_{k-2}}\wedge\cdots\wedge\tau_{Nn+i_{1}}\\
& =
\tau_{N(n-1)+i_{k-1}}\wedge\tau_{N(n-1)+i_{k-2}}\wedge\cdots\wedge\tau_{N(n-1)+i_{1}}
\\
& =\theta_{N(n-1)+i_k}.
\end{array} $$
The last equality holds true since by monotonicity we have
$\theta_{N(n-1)+i_k}\ge \theta_{Nn+i_k}$ and by definition
$$\theta_{N(n-1)+i_k}\le
\tau_{N(n-1)+i_{k-1}}\wedge\tau_{N(n-1)+i_{k-2}}\wedge\cdots\wedge
\tau_{N(n-1)+i_{1}}=\theta_{Nn+i_k}.$$ Therefore by definition of
$\tau_{N(n-1)+i_k}$ we have
$\tau_{Nn+i_k}=\tau_{N(n-1)+i_k}=\tau_{N(n-2)+i_k}$. Thus the
property is satisfied for $j=k$.

Assume now that the property is satisfied for $j=k,..,\ell+1$
($2\leq\ell+1\leq k$) and let us show it is also valid for $j=\ell$.
From (\ref{nul}) and (\ref{nul3}) we have
$$\tau_{Nn+i_\ell}=\tau_{N(n-1)+i_\ell}\ge \theta_{Nn+i_\ell}
$$
and
$$\ba{ll}
\theta_{Nn+i_\ell}&=
\tau_{Nn+i_{\ell-1}}\wedge\tau_{Nn+i_{\ell-2}}\wedge\cdots\wedge\tau_{Nn+i_{1}}
\wedge\tau_{Nn+i_{k}}
\wedge\tau_{Nn+i_{k-1}}\wedge\cdots\wedge\tau_{Nn+i_{\ell+1}}.\ea
$$
On the other hand $$\ba{ll}\theta_{N(n-1)+i_\ell}&=
\tau_{N(n-1)+i_\ell-1}\wedge\tau_{N(n-1)+i_\ell-2}\wedge\cdots\wedge\tau_{N(n-1)+i_\ell-N+1}\\
& =\tau_{Nn+i_{\ell-1}}\wedge...\wd
\tau_{Nn+i_{1}}\wedge\tau_{N(n-1)+i_{k}-N}\cdots\wedge\tau_{N(n-1)+i_{\ell+1}-N}.\ea
$$
This second equality is obtained in the same way as in (\ref{nul2})
in using (\ref{nul}) and the induction hypothesis. Therefore, once
more by the induction hypothesis, we have:
$$\ba{ll}\theta_{N(n-1)+i_\ell}&=
\tau_{Nn+i_{\ell-1}}\wedge...\wd
\tau_{Nn+i_{1}}\wedge\tau_{Nn+i_{k}}\cdots\wedge\tau_{Nn+i_{\ell+1}}\\
&=\theta_{Nn+i_\ell}.\ea
$$
It follows that
$$\tau_{Nn+i_\ell}=\tau_{N(n-1)+i_\ell}\ge \theta_{Nn+i_\ell}=\theta_{N(n-1)+i_\ell}
$$
and then $\tau_{N(n-1)+i_\ell}=\tau_{N(n-2)+i_\ell}$. Thus the
property is satisfied for $\ell$.
\medskip

Therefore for any $j\in \{1,...,k\}$ we have
$$
\tau_{Nn+i_j}=\tau_{N(n-1)+i_j}\ge
\theta_{Nn+i_j}=\theta_{N(n-1)+i_j}
$$
which implies that $A_n\subset A_{n-1}$, for any $n\geq 1$ and then
$P(A_n)=0$ for any $n\geq 0$ since $P(A_0)=0$. $\Box$ \bigskip

\no {\bf Step 2}: To proceed let $n\geq0$ and $\varepsilon>0$, then
we have:
$$\begin{array}{l} E [(Y^{i}_{T^*_{i}}-Q^{i}_{T^*_{i}})\;
\ind_{\{T^*_{i_1}=T^*_{i_2}=\cdots=T^*_{i_k}<R^*_{I}\}}]

\leq \\ \qq {\bf E}[ (Y^{i}_{T^*_{i}}-Q^{i}_{T^*_{i}})\;
\ind_{\bigcap_{j=1}^k\;
\{T^*_{i_1}=T^*_{i_2}=\cdots=T^*_{i_k}<R^*_{I};\;
R^*_{I}>\tau_{Nn+i_j}\geq\theta_{Nn+i_j}\}}]\,\,

+ \\\qq  \displaystyle\sum_{j=1}^k {\bf E}[
(Y^{i}_{T^*_{i}}-Q^{i}_{T^*_{i}})\; \ind_{
\{T^*_{i_1}=T^*_{i_2}=\cdots=T^*_{i_k}<R^*_{I};\; R^*_{I}\leq
\tau_{Nn+i_j}\}}]\,\,

+ \\\qq  \displaystyle\sum_{j=1}^k {\bf E}[
(Y^{i}_{T^*_{i}}-Q^{i}_{T^*_{i}})\; \ind_{
\{Y^{i}_{T^*_{i}}-Q^{i}_{T^*_{i}}\leq\frac{1}{\varepsilon};\;Y^{i_j}_{T^*_{i_j}}-X^{i_j}_{T^*_{i_j}}>\varepsilon
;\;T^*_{i_1}=T^*_{i_2}=\cdots=T^*_{i_k}<R^*_{I};\;
\tau_{Nn+i_j}<\theta_{Nn+i_j}\}}]\,\,

+ \\\qq  \displaystyle\sum_{j=1}^k {\bf E}[
(Y^{i}_{T^*_{i}}-Q^{i}_{T^*_{i}})\; \ind_{
\{Y^{i}_{T^*_{i}}-Q^{i}_{T^*_{i}}>\frac{1}{\varepsilon};
\;Y^{i_j}_{T^*_{i_j}}-X^{i_j}_{T^*_{i_j}}>\varepsilon
;\;T^*_{i_1}=T^*_{i_2}=\cdots=T^*_{i_k}<R^*_{I};\;
\tau_{Nn+i_j}<\theta_{Nn+i_j}\}}] \,\,+ \\ \qq
\displaystyle\sum_{j=1}^k {\bf E}[
(Y^{i}_{T^*_{i}}-Q^{i}_{T^*_{i}})\; \ind_{
\{Y^{i_j}_{T^*_{i_j}}-X^{i_j}_{T^*_{i_j}}\leq\varepsilon
;\;T^*_{i_1}=T^*_{i_2}=\cdots=T^*_{i_k}<R^*_{I};\;
\tau_{Nn+i_j}<\theta_{Nn+i_j}\}}].\end{array}
$$
Therefore, in taking into account (\ref{nul4}), we have:
$$\begin{array}{l}
E [(Y^{i}_{T^*_{i}}-Q^{i}_{T^*_{i}})\;
\ind_{\{T^*_{i_1}=T^*_{i_2}=\cdots=T^*_{i_k}<R^*_{I}\}}] \leq
 \\\qq  \displaystyle\sum_{j=1}^k {\bf E}[
(Y^{i}_{T^*_{i}}-Q^{i}_{T^*_{i}})\; \ind_{ \{T^*_{i_j}<R^*_{I};\;
R^*_{I}\leq \tau_{Nn+i_j}\}}]\,\, +
\\\qq \frac{1}{\varepsilon} \displaystyle\sum_{j=1}^k P
(Y^{i_j}_{T^*_{i_j}}-X^{i_j}_{T^*_{i_j}}>\varepsilon
;\;T^*_{i_j}=R^*_{i_j}<T;\; \tau_{Nn+i_j}<\theta_{Nn+i_j}) + \\
\qq k E [(Y^{i}_{T^*_{i}}-X^{i}_{T^*_{i}})\; \ind_{
\{Y^i_{T^*_{i}}-X^i_{T^*_{i}}>\frac{1}{\varepsilon} \}}]
 + \sum_{j=1}^k {\bf E}[
(Y^{i}_{T^*_{i}}-Q^{i}_{T^*_{i}})\; \ind_{
\{Y^{i_j}_{T^*_{i}}-X^{i_j}_{T^*_{i}}\leq\varepsilon , T^*_{i}<T
\}}].\ea
$$

Now taking first the limit as $n\rw \infty$ and using the second
property of Lemma \ref{lem3}-(ii) for the third term of the
right-hand side, then taking the limit as $\eps \rw 0$ to obtain:
$$\begin{array}{l} E [(Y^{i}_{T^*_{i}}-Q^{i}_{T^*_{i}})\;
\ind_{\{T^*_{i_1}=T^*_{i_2}=\cdots=T^*_{i_k}<R^*_{I}\}}] \leq
\displaystyle\sum_{j=1}^k {\bf E}[
(Y^{i}_{T^*_{i}}-Q^{i}_{T^*_{i}})\; \ind_{
\{Y^{i_j}_{i_j}-X^{i_j}_{T^*_i}=0  , T^*_{i}<T\}}]
\end{array} $$
But by Assumption (A4) we have ${\bf E}[
(Y^{i}_{T^*_{i}}-Q^{i}_{T^*_{i}})\; \ind_{
\{Y^{i_j}_{i_j}-X^{i_j}_{T^*_i}=0 , T^*_{i}<T\}}]=0$ therefore
$$ {\bf E}[ (Y^{i}_{T^*_{i}}-Q^{i}_{T^*_{i}})\;
\ind_{\{T^*_{i_1}=T^*_{i_2}=\cdots=T^*_{i_k}<R^*_{I}\}}]=0  \mbox{
and } E [(Y^i_{T^*_{i}}-Q^i_{T^*_{i}})\;
\ind_{\{T^*_{i}=R^*_{i}<T\}}]=0$$ which completes the proof of
(\ref{nul5}). \qed
\bigskip

We are now ready to give the main result of this paper which is a
direct consequence of Lemma \ref{lem3}-(ii), Proposition \ref{prop1}
and the fact that $Y^i\ge Q^i$ for any $i\in \cJ$.
\begin{thm} \label{existence} Under assumptions (A1)-(A4), the $N$-uplet of stopping times
$(T_i^*)_{i=1,...,N}$ is a Nash equilibrium point for the N-player
nonzero-sum Dynkin game associated with $(J_i)_{i\in \cJ}$.
\end{thm}

\begin{rem} Note that $(T^*_i)_{i\in \cJ}$ and $(J_i(T^*_1,T^*_2,\cdots,T^*_N ))_{i\in \cJ}$
do not depend on the processes $(Q^i_t)_{t<T}$ for any $i\in \cJ.$
On the other hand the NEP of a Dynkin game is not unique. Actually assume that for any $i\in \cJ$ and $t\leq T$, $X^i_t=\frac{1}{2}$ and 
$Q^i_t=Y^i_t=1$. Therefore one can easily show, directly or in using Proposition \ref{streamline}, that 
for any $t_0\in [0,T]$, $(t_0,...,t_0)$ is a NEP for this game and $J_i(t_0,...,t_0)=1$, $\forall i\in \cJ$. \qed
\end{rem}

Finally we have the following result related to Proposition
\ref{streamline} which is a direct consequence of the fact that
$(T_i^*)_{i\in \cJ}$  is a NEP for the nonzero-sum Dynkin game  and,
Proposition \ref{snev} and Remark \ref{snelmart} and Proposition
\ref{prop1}.

\begin{prop}For $i\in \cJ$, let us set:
$$
W^i:=(W^i_t)_{t\leq T}:={\bf R}(X^i_t\ind_{\{t<R_i^*\}}+\tilde
Y^i_{R_i^*}\ind_{\{t\geq R_i^*\}})
$$ where ${\bf R}$ is the Snell envelope operator.
Then $(T_i^*)_{i\in \cJ}$ and  $(W^1,...,W^N)$ satisfy (i)-(iii) of
Proposition \ref{streamline}. \qed

\end{prop}

\end{document}